\documentclass[a4paper]{article}
\usepackage[latin1]{inputenc} 
\usepackage{color}
\usepackage{amsthm}
\usepackage{amsmath}
\usepackage{graphicx}
\usepackage{amssymb}
\usepackage{bbold}
\usepackage{esint}
\usepackage{subfigure}
\usepackage{color}

\usepackage{anysize}
\marginsize{2cm}{2cm}{2cm}{2cm}

\def\beq{\begin{equation}}
\def\eeq{\end{equation}}
\def\bea{\begin{eqnarray}}
\def\eea{\end{eqnarray}}
\def\nn{\nonumber}

\makeatletter

\theoremstyle{plain}

  \theoremstyle{remark}

\title{Non-Abelian Fusion Rules from an Abelian System}

\author{Pramod Padmanabhan\footnote{pramod23phys@gmail.com},  Paulo Teotonio-Sobrinho \footnote{teotonio@if.usp.br}, \\ \\
\small{Departmento de F\'{i}sica Matem\'{a}tica Universidade de S\~{a}o Paulo - USP,} \\
\small{CEP 05508-090 Cidade Universit\'aria, S\~{a}o Paulo - Brasil}}

\date{\today}
\begin{document}

\maketitle

\begin{abstract}

We demonstrate the emergence of non-Abelian fusion rules for excitations of a two dimensional lattice model built out of Abelian degrees of freedom. It can be considered as an extension of the usual toric code model on a two dimensional lattice augmented with matter fields. It consists of the usual $\mathbb{C}(\mathbb{Z}_p)$ gauge degrees of freedom living on the links together with matter degrees of freedom living on the vertices. The matter part is described by a $n$ dimensional vector space which we call $H_n$. The $\mathbb{Z}_p$ gauge particles act on the vertex particles and thus $H_n$ can be thought of as a $\mathbb{C}(\mathbb{Z}_p)$ module. An exactly solvable model is built with operators acting in the corresponding Hilbert space. The vertex excitations for this model are studied and shown to obey non-Abelian fusion rules. We will show this for specific values of $n$ and $p$, though we believe this feature holds for all $n>p$. We will see that non-Abelian anyons of the quantum double of $\mathbb{C}(S_3)$ are obtained as part of the vertex excitations of the model with $n=6$ and $p=3$. Ising anyons are obtained in the model with $n=4$ and $p=2$. The $n=3$ and $p=2$ case is also worked out as this is the simplest model exhibiting non-Abelian fusion rules. Another common feature shared by these models is that the ground states have a higher symmetry than $\mathbb{Z}_p$. This makes them possible candidates for realizing quantum computation. 

\end{abstract} 

\section{Introduction}

The concept of topological order has come a long way since its inception to describe phenomena such as high temperature superconductivity and the fractional quantum Hall effect(FQHE)~\cite{WenBook, DSP, PG, PWA}. It is now believed that topological order is classified into two broad types: symmetry protected topological phases(SPT)~\cite{XChenSPT} described by short-ranged entangled states (those which can be transformed to product states using local unitary transformations(LUTs)) and phases with intrinsic topological order which are described by long-ranged entangled states (those which cannot be transformed to product states using LUTs). The SPT states have no fractional excitations in the bulk and typically have edge states with well known examples being topological insulators~\cite{TIReview, TIReview2}. In contrast to this, long-ranged entangled states have fractional excitations in the bulk and a non-zero entanglement entropy~\cite{setPhase}. In two dimensions these systems have anyons as part of their low energy excitations~\cite{Ref1, Ref2, Ref3, Ref4}. The prototype for such models is given by the quantum double models of Kitaev~\cite{KitToric, AguadoQDM}.

Anyonic particle in two dimensions satisfy exotic statistics with the Abelian versions having phases and non-Abelian versions, unitary matrices as their statistics parameters. The non-Abelian anyons have found application in fault-tolerant quantum computation~\cite{KitToric, NR, PreskillNotes, Ref8}. There are several ways to realize these particles in actual physical systems~\cite{Ref11, Ref12, Ref13, Ref14}. It is interesting to find lattice models in two dimensions where we can find such particles~\cite{KitToric, AguadoQDM, LW, Ref17, Ref18}. 

Given the difficulty in realizing non-Abelian anyons in the lab it is of interest to find Abelian systems which have non-Abelian particles in their spectrum. Several efforts have been made in this direction with Kitaev showing that his Honeycomb model~\cite{KitHoney} with magnetic perturbations support Ising anyons. In another approach Bombin uses defects in lattices to show the existence of non-Abelian anyons from the toric code~\cite{BombDefect}. This work has been extended in~\cite{StanfordGroup}. Wooton et al engineer non-Abelian anyons from Abelian models by using the superposition of states from the Abelian system~\cite{PachosIsing, PachosIsing2}. 

Taking this program forward we introduce a class of Abelian models in the two dimensional lattice which contain non-Abelian excitations. The systems we consider have gauge and matter degrees of freedom living on the links and the vertices of the two dimensional lattice respectively. The gauge fields belong to the group algebra of discrete Abelian gauge groups namely $\mathbb{C}(\mathbb{Z}_p)$ and the matter fields form a module for this algebra which we call $H_n$. In other words $H_n$ is a $n$-dimensional vector space with an inner product and it carries a $n$-dimensional representation of the gauge group algebra $\mathbb{C}(\mathbb{Z}_p)$ which means that the gauge group algebra acts on this vector space. We only consider the case with $n>p$ and study three particular examples of $(n,p)$ namely $(3,2), (4,2)$ and $(6,3)$. We will show the construction of exactly solvable models using these gauge and matter fields as our input data. These models are similar to the quantum double models and can be thought of as quantum double models augmented with matter fields on the vertices. We denote these models $H_n/\mathbb{C}(\mathbb{Z}_p)$. We will study vertex excitations in this model and show them to obey non-Abelian fusion rules. 

The paper is organized as follows. Section 2 describes the ingredients of the exactly solvable model. These include the vertex and link degrees of freedom which fix the Hilbert space and the operators which are used to build the Hamiltonian of the theory. We then go on to describe vertex excitations of this system by constructing excitation operators in section 3. The excitations are worked out for three pairs of $(n,p)$ namely $(3,2)$, $(4,2)$ and $(6,3)$. Section 4 describes the fusion rules of these excitations which help us identify the anyons in the cases where they have already been studied. We conclude with a few remarks in section 5. 

\section{The Model}

The system we consider can live on an arbitrary two dimensional lattice with degrees of freedom on the vertices and the links. For simplicity we will work on the square lattice. The gauge degrees of freedom living on the links belong the group algebra of the gauge group, namely $\mathbb{C}(\mathbb{Z}_p)$. The matter degrees of freedom belong to a module of this group algebra and we denote it by the n-dimensional vector space $H_n$. We consider the case where $n>p$ to obtain useful modules for the gauge group algebra. 

The basis elements of $\mathbb{C}(\mathbb{Z}_p)$ is given by $\{ \phi_{\omega^k} | k\in (0, 1, \cdots p-1)\}$ where $\omega = e^{\frac{2\pi i}{p}}$. For each link $l$ we have a Hilbert space $\mathcal{H}_l$ with basis $\{\phi_{\omega^k}\}$. Consider the module to be made up of the basis elements $H_n = \{\chi_1, \cdots, \chi_n\}$. The inner product in this module is $\langle\chi_i|\chi_j\rangle = \delta_{ij}$. For each vertex $v$ we have a Hilbert space $\mathcal{H}_v$ with basis $\{\chi_j\}$. The full Hilbert space is given by $$\mathcal{H}=\bigotimes_l\mathcal{H}_l\bigotimes_v\mathcal{H}_v.$$ The total number of tensor products equal $N_l+N_v$ where $N_l$ and $N_v$ are the total number of links and vertices respectively. The dimension of $\mathcal{H}$ equals $p^{N_l}\times n^{N_v}$. 

The action of the basis elements of $\mathbb{C}(\mathbb{Z}_p)$ on the module $H_n$ is given by the representation matrices $\mu(\phi_{\omega^k})$. The action is fixed once we define $\mu(\phi_{\omega})$. $\mu(\phi_1)$ is the $n$ by $n$ identity matrix. The action of $\phi_{\omega}$ is obtained by permuting $p$ of the $n$ basis elements of the module leaving the remaining $n-p$ basis elements invariant. We can consider unitarily equivalent modules obtained by the unitary transformations generated by $X$ and $Z$ which are the generators of $\mathbb{Z}_p$. They are given by 
\bea\label{x} X.\phi_{\omega^k} & = & \phi_{\omega^{k+1}} \nn \\
 Z.\phi_{\omega^k} & = & \omega^k \phi_{\omega^k}.\eea

The operators making up the Hamiltonian are given by vertex operators, $A_v$ and link operators, $C_l$. The vertex operators can be thought of as gauge transformations and they are defined as follows

\beq A_v = \frac{1}{p}\sum_{k=0}^{p-1}~\left[\mu_v(\phi_{\omega^k})\otimes X_{i_1}^k \otimes X_{i_2}^k \otimes X_{i_3}^{-k} \otimes X_{i_4}^{-k}\right] \eeq
where $X$ is given by Eq.(\ref{x}). 

The link operator acts on a link, $l$ and two adjacent vertices $v_1$ and $v_2$. It is defined as
\beq\label{link} C_l |\chi_{v_1}, \chi_{v_2}, \phi_l\rangle = \langle\mu(\phi_l).\chi_{v_1}|\chi_{v_2}\rangle~|\chi_{v_1}, \chi_{v_2}, \phi_l\rangle \eeq
It is possible to obtain a matrix representation of $C_l$ and we will do this in the sections where we work with specific examples.

\begin{figure}
	      \begin{center}
		\includegraphics[scale=1.25]{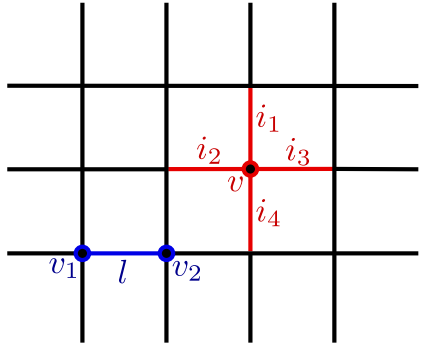}
		\caption{The action of the vertex operator $A_v$ is shown in red and the action of the link operator $C_l$ is shown in blue.}
		\label{figshort}
		\end{center}

\end{figure}

The action of these operators is shown in fig(\ref{figshort}).

It is easy to show that both $A_v$ and $C_l$ are projectors and they commute with each other. This implies that a Hamiltonian formed out of these operators will be exactly solvable. An example of such an Hamiltonian is given by
\beq H = \alpha\sum_v A_v + \beta\sum_l C_l \eeq 
where $\alpha$ and $\beta$ are parameters. 

We can also define a plaquette operator, $B_p$, similar to the ones found in quantum double models~\cite{KitToric, AguadoQDM} in this model. This operator will only act on the gauge fields on the links and can be considered to act trivially on the matter fields on the vertices. Since we are only concerned with vertex excitations in this paper we do not consider the effect of these plaquette operators in this model. 

We will now consider examples of the $H_n/\mathbb{C}(\mathbb{Z}_p)$ model for specific values of $(n,p)$. 

\subsection{$(n, p) = (3,2)$:}

The vertex operator in this theory is given by
\beq\label{star1} A_v = \frac{\mathbb{1}\otimes\mathbb{1}\otimes\mathbb{1}\otimes\mathbb{1}\otimes\mathbb{1} + \mu_v(\phi_{-1})\otimes\sigma^x_{i_1}\otimes\sigma^x_{i_2}\otimes\sigma^x_{i_3}\otimes\sigma^x_{i_4}}{2} \eeq 
where $\sigma^x$ is the Pauli matrix and indices $v$, $i_1$, $i_2$, $i_3$ and $i_4$ are shown in fig(\ref{figshort}). The orthogonal operators become important in finding the excitations and hence we will also write them down. The operator orthogonal to Eq.(\ref{star1}) is given by 
\beq\label{star2} A_v^\perp = \frac{\mathbb{1}\otimes\mathbb{1}\otimes\mathbb{1}\otimes\mathbb{1}\otimes\mathbb{1} - \mu_v(\phi_{-1})\otimes\sigma^x_{i_1}\otimes\sigma^x_{i_2}\otimes\sigma^x_{i_3}\otimes\sigma^x_{i_4}}{2}. \eeq

The operator $\mu_v(\phi_{-1})$ is given by the following matrix
\beq\label{mu32} \mu_v(\phi_{-1}) = \left(\begin{array}{ccc} 0 & 1 & 0 \\ 1 & 0 & 0 \\ 0 & 0 & 1\end{array}\right).\eeq 

The link operator $C_l$ is given by 
\begin{eqnarray}\label{link32} C_l & = & \left[\frac{\mathbb{1} + Z_{v_1} + Z_{v_1}^2}{3}\right]\otimes\left[\frac{1+\sigma^z_l}{2}\right]\otimes\left[\frac{\mathbb{1} + Z_{v_2} + Z_{v_2}^2}{3}\right] \nn \\
                                  & + & \left[\frac{\mathbb{1} + \omega^2Z_{v_1} + \omega Z_{v_1}^2}{3}\right]\otimes\left[\frac{1-\sigma^z_l}{2}\right]\otimes\left[\frac{\mathbb{1} + Z_{v_2} + Z_{v_2}^2}{3}\right] \nn \\
                                  & + & \left[\frac{\mathbb{1} + Z_{v_1} + Z_{v_1}^2}{3}\right]\otimes\left[\frac{1-\sigma^z_l}{2}\right]\otimes\left[\frac{\mathbb{1} + \omega^2Z_{v_2} + \omega Z_{v_2}^2}{3}\right] \nn \\ 
                                  & + & \left[\frac{\mathbb{1} + \omega^2Z_{v_1} + \omega Z_{v_1}^2}{3}\right]\otimes\left[\frac{1+\sigma^z_l}{2}\right]\otimes\left[\frac{\mathbb{1} +\omega^2 Z_{v_2} +\omega Z_{v_2}^2}{3}\right] \nn \\
                                  & + & \left[\frac{\mathbb{1} + \omega Z_{v_1} + \omega^2Z_{v_1}^2}{3}\right]\otimes\mathbb{1}\otimes\left[\frac{\mathbb{1} + \omega Z_{v_2} + \omega^2Z_{v_2}^2}{3}\right] \end{eqnarray} 
where $Z$ is a generator of $\mathbb{Z}_3$ given by 
\beq\label{z4} Z = \left(\begin{array}{ccc} 1 & 0 & 0 \\ 0 & \omega & 0 \\ 0 & 0 & \omega^2 \end{array}\right) \eeq
and $\omega = e^{i\frac{2\pi}{3}}$ and indices $v_1$, $v_2$ and $l$ are shown in fig(\ref{figshort}).  

\subsection{$(n, p) = (4,2)$:}

The vertex operator in this theory is given by
\beq\label{star3} A_v = \frac{\mathbb{1}\otimes\mathbb{1}\otimes\mathbb{1}\otimes\mathbb{1}\otimes\mathbb{1} + \mu_v(\phi_{-1})\otimes\sigma^x_{i_1}\otimes\sigma^x_{i_2}\otimes\sigma^x_{i_3}\otimes\sigma^x_{i_4}}{2} \eeq 
where $\sigma^x$ is the Pauli matrix. The operator orthogonal to this is given by 
\beq\label{star4} A_v^\perp = \frac{\mathbb{1}\otimes\mathbb{1}\otimes\mathbb{1}\otimes\mathbb{1}\otimes\mathbb{1} - \mu_v(\phi_{-1})\otimes\sigma^x_{i_1}\otimes\sigma^x_{i_2}\otimes\sigma^x_{i_3}\otimes\sigma^x_{i_4}}{2}. \eeq

The operator $\mu_v(\phi_{-1})$ in the this case is given by the following matrix

\beq\label{mu42} \mu_v(\phi_{-1}) = \left(\begin{array}{cc} \sigma^x & 0 \\ 0 & \sigma^x\end{array}\right).\eeq

The link operator $C_l$ in this case is given by

\bea\label{link42} C_l & = & \left[\frac{\mathbb{1}+Z_{v_1}+Z_{v_1}^2+Z_{v_1}^3}{4}\right]\otimes\left[\frac{\mathbb{1}+\sigma^z_l}{2}\right]\otimes\left[\frac{\mathbb{1}+Z_{v_2}+Z_{v_2}^2+Z_{v_2}^3}{4}\right] \nn \\
                                   & + & \left[\frac{\mathbb{1}+\omega^3Z_{v_1}+\omega^2Z_{v_1}^2+\omega Z_{v_1}^3}{4}\right]\otimes\left[\frac{\mathbb{1}-\sigma^z_l}{2}\right]\otimes\left[\frac{\mathbb{1}+Z_{v_2}+Z_{v_2}^2+Z_{v_2}^3}{4}\right] \nn \\
                                   & + & \left[\frac{\mathbb{1}+Z_{v_1}+Z_{v_1}^2+Z_{v_1}^3}{4}\right]\otimes\left[\frac{\mathbb{1}-\sigma^z_l}{2}\right]\otimes\left[\frac{\mathbb{1}+\omega^3Z_{v_2}+\omega^2Z_{v_2}^2+\omega Z_{v_2}^3}{4}\right] \nn \\
                                   & + & \left[\frac{\mathbb{1}+\omega^3Z_{v_1}+\omega^2Z_{v_1}^2+\omega Z_{v_1}^3}{4}\right]\otimes\left[\frac{\mathbb{1}+\sigma^z_l}{2}\right]\otimes\left[\frac{\mathbb{1}+\omega^3Z_{v_2}+\omega^2Z_{v_2}^2+\omega Z_{v_2}^3}{4}\right] \nn \\
                                   & + & \left[\frac{\mathbb{1}+\omega^2Z_{v_1}+Z_{v_1}^2+\omega^2Z_{v_1}^3}{4}\right]\otimes\left[\frac{\mathbb{1}+\sigma^z_l}{2}\right]\otimes\left[\frac{\mathbb{1}+\omega^2Z_{v_2}+Z_{v_2}^2+\omega^2Z_{v_2}^3}{4}\right] \nn \\
                                   & + & \left[\frac{\mathbb{1}+\omega Z_{v_1}+\omega^2Z_{v_1}^2+\omega^3Z_{v_1}^3}{4}\right]\otimes\left[\frac{\mathbb{1}-\sigma^z_l}{2}\right]\otimes\left[\frac{\mathbb{1}+\omega^2Z_{v_2}+Z_{v_2}^2+\omega^2Z_{v_2}^3}{4}\right] \nn \\
                                   & + & \left[\frac{\mathbb{1}+\omega^2Z_{v_1}+Z_{v_1}^2+\omega^2Z_{v_1}^3}{4}\right]\otimes\left[\frac{\mathbb{1}-\sigma^z_l}{2}\right]\otimes\left[\frac{\mathbb{1}+\omega Z_{v_2}+\omega^2Z_{v_2}^2+\omega^3Z_{v_2}^3}{4}\right] \nn \\
                                   & + & \left[\frac{\mathbb{1}+\omega Z_{v_1}+\omega^2Z_{v_1}^2+\omega^3Z_{v_1}^3}{4}\right]\otimes\left[\frac{\mathbb{1}+\sigma^z_l}{2}\right]\otimes\left[\frac{\mathbb{1}+\omega Z_{v_2}+\omega^2Z_{v_2}^2+\omega^3Z_{v_2}^3}{4}\right] \eea
 where $Z$ is a generator of $\mathbb{Z}_4$ given by 
\beq\label{z3} Z = \left(\begin{array}{cccc} 1 & 0 & 0 & 0 \\ 0 & \omega & 0 & 0 \\ 0 & 0 & \omega^2 & 0 \\ 0 & 0 & 0 & \omega^3 \end{array}\right) \eeq
and $\omega = e^{i\frac{2\pi}{4}}$.

\subsection{$(n,p) = (6,3)$:}

 The vertex operator in this theory is given by
 \beq\label{star5} A_v^1 = \frac{\mathbb{1}\otimes\mathbb{1}\otimes\mathbb{1}\otimes\mathbb{1}\otimes\mathbb{1}+ \mu_v(\phi_\omega)\otimes X_{i_1}\otimes X_{i_2}\otimes X^2_{i_3}\otimes X^2_{i_4} + \mu_v(\phi_{\omega^2})\otimes X^2_{i_1}\otimes X^2_{i_2}\otimes X_{i_3}\otimes X_{i_4}}{3} \eeq
 where $X$ is a generator of $\mathbb{Z}_3$ given by
 \beq\label{x3} X = \left(\begin{array}{ccc} 0 & 0 & 1\\ 1 & 0 & 0 \\ 0 & 1 & 0\end{array}\right)\eeq 
 and $\omega = e^{\frac{2\pi i}{3}}$.
 
There are two orthogonal operators to this vertex operator given by
 \beq\label{star6} A_v^2 = \frac{\mathbb{1}\otimes\mathbb{1}\otimes\mathbb{1}\otimes\mathbb{1}\otimes\mathbb{1}+ \omega^2\mu_v(\phi_\omega)\otimes X_{i_1}\otimes X_{i_2}\otimes X^2_{i_3}\otimes X^2_{i_4} + \omega \mu_v(\phi_{\omega^2})\otimes X^2_{i_1}\otimes X^2_{i_2}\otimes X_{i_3}\otimes X_{i_4}}{3} \eeq
and
 \beq\label{star7} A_v^3 = \frac{\mathbb{1}\otimes\mathbb{1}\otimes\mathbb{1}\otimes\mathbb{1}\otimes\mathbb{1}+ \omega \mu_v(\phi_\omega)\otimes X_{i_1}\otimes X_{i_2}\otimes X^2_{i_3}\otimes X^2_{i_4} + \omega^2\mu_v(\phi_{\omega^2})\otimes X^2_{i_1}\otimes X^2_{i_2}\otimes X_{i_3}\otimes X_{i_4}}{3}. \eeq
 
 The expression for $\mu_v(\phi_\omega)$ is given by
 \beq \mu_v(\phi_\omega)  =  \left(\begin{array}{cc} X & 0 \\ 0 & X\end{array}\right).\eeq
 
 The expression for the link operator $C_l$ in this case is more complicated than the previous two cases. However there is a systematic way of writing down this operator which we explain now. This method also explains how we obtained the expressions for the link operators of the previous two cases. 
 
 From the definition of the link operator in Eq.(\ref{link}) we see that it is diagonal and the proportionality factor is given by the inner product in Eq.(\ref{link}). 
We can insert a complete set of states into this inner product to obtain
\beq \langle\mu(\phi_l).\chi_{v_1}|\chi_{v_2}\rangle = \sum_v\langle \mu(\phi_l).\chi_{v_1}|\chi_v\rangle\langle\chi_v|\chi_{v_2}\rangle. \eeq
The expression $\langle\chi_v|\chi_{v_2}\rangle$ is the eigenvalue of a projector to one of the basis elements of the module. If the module we are working with is $H_n$ then the projector having this inner product as the eigenvalue can be written in terms of $Z$, the generators of $\mathbb{Z}_n$. 
The expression $\langle\mu(\phi_l).\chi_{v_1}|\chi_v\rangle$ is obtained by projecting to all the configurations $\phi_l$ and $\chi_{v_1}$ which gets projected to $\chi_v$. These can again be written in terms of the generators of $\mathbb{Z}_n$ and $\mathbb{Z}_p$ respectively. 

Using this procedure we can obtain the link operator for the $(6,3)$ case. 

This procedure also helps us obtain the orthogonal operators to the link operators we wrote down in the three cases. These orthogonal operators are given by 
\beq\label{linkortho} C_l^k |\chi_{v_1}, \chi_{v_2}, \phi_l\rangle = \langle\mu(\phi_l).\chi_{v_1}|X^k|\chi_{v_2}\rangle~|\chi_{v_1}, \chi_{v_2}, \phi_l\rangle \eeq
where $k\in(1,\cdots n-1)$ and $X$ is the generator of $\mathbb{Z}_n$. This gives us $n-1$ orthogonal projectors to the link operators written previously for matter fields living in a module $H_n$.  The matrix representation for these operators are found in the same way as was found for the link operators in Eq.(\ref{link}). 

The orthogonal operators become important in finding the excitations for these models. In the following section we will see how this happens.

\section{Vertex Excitations in the $H_n/\mathbb{C}(\mathbb{Z}_p)$ Models}

In what follows we will consider vertex excitations, that is the excitations which arise by acting on just the vertices. In order to obtain these excitations we need to find operators which will help us transform a given star and link operator to its orthogonal operator. This can be understood by looking at the $\mathbb{Z}_2$-toric code example. Consider the star operators in this theory. They are projectors and have eigenvalues 1 and 0. Let us assume the coefficients of the star operators in the Hamiltonian to be such that the ground state is an eigenstate of this operator when its eigenvalue is 1. This means that excitations correspond to the eigenvalue 0. So the operator that does this job in the toric code case is $\sigma^z_l$ acting on a link $l$. That is 
\beq A_v\sigma^z_l|G\rangle = \sigma^z_lA_v^\perp|G\rangle \eeq
where $A_v|G\rangle = |G\rangle$ is one of the ground state conditions for the $\mathbb{Z}_2$-toric code.

But $A_v^\perp|G\rangle = 0$ as $A_vA_v^\perp = 0$. Thus $\sigma^z_l$ is the operator we are looking for that transforms $A_v$ to its orthogonal counterpart thereby creating excitations of $A_v$. These are none other than the charge excitations of the $\mathbb{Z}_2$-toric code. 

We will now obtain such operators which create the vertex excitations for the three cases of $(n,p)$ considered in the previous section. 

 \subsection{$(n,p)=(3,2)$:}
 
 Consider the following set of operators 
\begin{eqnarray}\label{excitations32} X_{12} & = & \left(\begin{array}{ccc} 0 & 1 & 0 \\ 1 & 0 & 0 \\ 0 & 0 & 0\end{array}\right) ~;~ X_{23} = \left(\begin{array}{ccc} 0 & 0 & 0 \\ 0 & 0 & 1 \\ 0 & 1 & 0\end{array}\right)~;~ X_{31} = \left(\begin{array}{ccc} 0 & 0 & 1 \\ 0 & 0 & 0 \\ 1 & 0 & 0\end{array}\right), \nn \\ 
Z_{12} & = & \left(\begin{array}{ccc} 1 & 0 & 0 \\ 0 & -1 & 0 \\ 0 & 0 & 0\end{array}\right) ~;~ Z_{23} = \left(\begin{array}{ccc} 0 & 0 & 0 \\ 0 & 1 & 0 \\ 0 & 0 & -1\end{array}\right)~;~ Z_{31} = \left(\begin{array}{ccc} 1 & 0 & 0 \\ 0 & 0 & 0 \\ 0 & 0 & -1\end{array}\right), \nn \\ 
Y_{12} & = & \left(\begin{array}{ccc} 0 & -i & 0 \\ i & 0 & 0 \\ 0 & 0 & 0\end{array}\right) ~;~ Y_{23} = \left(\begin{array}{ccc} 0 & 0 & 0 \\ 0 & 0 & -i \\ 0 & i & 0\end{array}\right)~;~ Y_{31} = \left(\begin{array}{ccc} 0 & 0 & -i \\ 0 & 0 & 0 \\ i & 0 & 0\end{array}\right). \end{eqnarray} 

It can be seen that these matrices span the vector space of three by three matrices by recalling the fact that $\mathbb{1}$, $X$, $X^2$, $Z$, $Z^2$, $XZ$, $XZ^2$, $X^2Z$, $X^2Z^2$ span the set of three by three matrices and
\beq X = \frac{1}{2}\left[X_{12} -iY_{12} + X_{23} - iY_{23} + X_{31} + iY_{31}\right] \eeq 
\beq Z = \omega Z_{23} + Z_{13}\eeq
where $X$ and $Z$ are given by Eq.(\ref{x3}) and Eq.(\ref{z4}) respectively. 

If we consider the ground state to be $|G\rangle$ then $X_{12}|G\rangle$ is an eigenstate of the star operator with eigenvalue 1 and an eigenstate of the link operator with eigenvalue 0. Similarly $Z_{12}|G\rangle$ is an eigenstate of the link operator with eigenvalue 1 and an eigenstate of the star operator with eigenvalue 0. 

Now we need to consider the action of the other operators namely $X_{23}$, $X_{31}$, $Z_{23}$ etc. Consider $A_vO|G\rangle = aO|G\rangle$ and $C_lO|G\rangle = bO|G\rangle$ with $A_v$ and $C_l$ given by Eq.(\ref{star1}) and Eq.(\ref{link32}) respectively. We will write down the operators $O$, and the values of $a$ and $b$ in the form of a table(\ref{sometable}). 

\begin{table}
\begin{center}
\begin{tabular}{|c|c|c|}
\hline
$O$ & $a$ & $b$ \\
\hline
$Q_-^z =Z_{31}-Z_{23}= Z_{12}$ & 0 & 1 \\ 
$X_{12}$ & 1 & 0 \\
$Y_{12}$ & 0 & 0 \\
$Q_+^z=Z_{23}+Z_{31}$ & 1 & 1 \\
$Q_-^x = X_{23}-X_{31}$ & 0 & 0 \\
$Q_+^x = X_{23}+X_{31}$ & 1 & 0 \\
$Q_+^y= Y_{23}+Y_{31}$ & 1 & 0 \\
$Q_-^y= Y_{23}-Y_{31}$ & 0 & 0 \\
\hline
\end{tabular}
\end{center}
\caption{Operators creating the vertex excitations} \label{sometable}
\end{table}

The following operator acting on a vertex $v$ commutes with $A_v$ and $C_l$
\beq R_v = \left(\begin{array}{ccc} a & 0 & 0 \\ 0 & a & 0 \\ 0 & 0 & b\end{array}\right) \eeq
where $a$ and $b$ are two arbitrary constants. This generates a bigger symmetry than $\mathbb{Z}_2$ for the ground states of the theory. More precisely it is a continuous local symmetry. The excited states however do not have this symmetry as this operator does not commute with the excitations in Eq.(\ref{excitations32}). Thus the excited states have a degeneracy different from that of the ground states. We will elaborate more on this later. 

 \subsection{$(n,p)=(4,2)$:}
 
 The operators which form a basis of the vector space of four by four matrices are given by
 \bea\label{excitations42} X_{12} & = & \left(\begin{array}{cccc} 0 & 1 & 0 & 0 \\ 1 & 0 & 0 & 0 \\ 0 & 0 & 0 & 0 \\ 0 & 0 & 0 & 0\end{array}\right);~X_{13} = \left(\begin{array}{cccc} 0 & 0 & 1 & 0 \\ 0 & 0 & 0 & 0 \\ 1 & 0 & 0 & 0 \\ 0 & 0 & 0 & 0\end{array}\right);~X_{14} = \left(\begin{array}{cccc} 0 & 0 & 0 & 1 \\ 0 & 0 & 0 & 0 \\ 0 & 0 & 0 & 0 \\ 1 & 0 & 0 & 0\end{array}\right); \nn \\ X_{23} & = & \left(\begin{array}{cccc} 0 & 0 & 0 & 0 \\ 0 & 0 & 1 & 0 \\ 0 & 1 & 0 & 0 \\ 0 & 0 & 0 & 0\end{array}\right);~X_{24} = \left(\begin{array}{cccc} 0 & 0 & 0 & 0 \\ 0 & 0 & 0 & 1 \\ 0 & 0 & 0 & 0 \\ 0 & 1 & 0 & 0\end{array}\right);~X_{34} = \left(\begin{array}{cccc} 0 & 0 & 0 & 0 \\ 0 & 0 & 0 & 0 \\ 0 & 0 & 0 & 1 \\ 0 & 0 & 1 & 0\end{array}\right), \nn \\ 
 Y_{12} & = & \left(\begin{array}{cccc} 0 & -i & 0 & 0 \\ i & 0 & 0 & 0 \\ 0 & 0 & 0 & 0 \\ 0 & 0 & 0 & 0\end{array}\right);~Y_{13} = \left(\begin{array}{cccc} 0 & 0 & -i & 0 \\ 0 & 0 & 0 & 0 \\ i & 0 & 0 & 0 \\ 0 & 0 & 0 & 0\end{array}\right);~Y_{14} = \left(\begin{array}{cccc} 0 & 0 & 0 & -i \\ 0 & 0 & 0 & 0 \\ 0 & 0 & 0 & 0 \\ i & 0 & 0 & 0\end{array}\right); \nn \\ Y_{23} & = & \left(\begin{array}{cccc} 0 & 0 & 0 & 0 \\ 0 & 0 & -i & 0 \\ 0 & i & 0 & 0 \\ 0 & 0 & 0 & 0\end{array}\right);~Y_{24} = \left(\begin{array}{cccc} 0 & 0 & 0 & 0 \\ 0 & 0 & 0 & -i \\ 0 & 0 & 0 & 0 \\ 0 & i & 0 & 0\end{array}\right);~Y_{34} = \left(\begin{array}{cccc} 0 & 0 & 0 & 0 \\ 0 & 0 & 0 & 0 \\ 0 & 0 & 0 & -i \\ 0 & 0 & i & 0\end{array}\right), \nn \\
  Z_{13} & = & \left(\begin{array}{cccc}1 & 0 & 0 & 0 \\ 0 & 0 & 0 & 0 \\ 0 & 0 & -1 & 0 \\ 0 & 0 & 0 & 0\end{array}\right);~ Z_{14} = \left(\begin{array}{cccc}1 & 0 & 0 & 0 \\ 0 & 0 & 0 & 0 \\ 0 & 0 & 0 & 0 \\ 0 & 0 & 0 & -1\end{array}\right) \nn \\                                                                                                                                                                                                                                                                                                                                                                        
  Z_{24} & = & \left(\begin{array}{cccc}0 & 0 & 0 & 0 \\ 0 & 1 & 0 & 0 \\ 0 & 0 & 0 & 0 \\ 0 & 0 & 0 & -1\end{array}\right);~ Z_{23} = \left(\begin{array}{cccc}0 & 0 & 0 & 0 \\ 0 & 1 & 0 & 0 \\ 0 & 0 & -1 & 0 \\ 0 & 0 & 0 & 0\end{array}\right) \eea                                  

It is possible to write down a table similar to table(\ref{sometable}) as in the $(3,2)$ case. Instead of doing this we will just look at those excitations which mimic the Ising anyons which obey non-Abelian fusion rules. These operators are given by
\beq\label{ising} \Sigma_k = \left(\begin{array}{cc} \sigma^x & 0 \\ 0 & \sigma^x\end{array}\right) +  k\left(\begin{array}{cc} \sigma^z & 0 \\ 0 & \sigma^z\end{array}\right) \eeq
where $k$ is an arbitrary constant. A similar construction was used to construct Ising anyons by superposing excitations of the $\mathbb{Z}_2$-toric code in~\cite{PachosIsing}. 

The fusion rules will be shown to be that of Ising anyons in the next section.  

As in the $(3,2)$ case there is an operator acting on the vertex that commutes with both the vertex and link operators. This is given by
\beq R_v = \left(\begin{array}{cc} a\mathbb{1} & 0 \\ 0 & b\mathbb{1}\end{array}\right)\eeq 
where $a$ and $b$ are arbitrary constants and $\mathbb{1}$ is the two by two identity matrix. Clearly not all the operators in Eq.(\ref{excitations42}) creating excitations commute with this operator thereby having the same effect on the degeneracy of the excited states when compared to the ground states as in the $(3,2)$ case discussed above. 

\subsection{$(n,p)=(6,3)$:}

Consider the six by six matrices to be made up of block matrices of two three by three matrices. Since we already know the basis of the three by three matrices from the previous model we can readily write down the basis of the excitations of the six by six matrices. These are given by 
\beq\label{excitations63} \left[\left(\begin{array}{cc} A & 0 \\ 0 & 0\end{array}\right) , \left(\begin{array}{cc} 0 & A \\ 0 & 0\end{array}\right), \left(\begin{array}{cc} 0 & 0 \\ A & 0\end{array}\right), \left(\begin{array}{cc} 0 & 0 \\ 0 & A\end{array}\right)\right] \eeq
where 
\beq A\in \{\mathbb{1}, X, X^2, Z, Z^2, XZ, XZ^2, X^2Z, X^2Z^2\} \eeq
where $X$ and $Z$ are given by Eq.(\ref{x3}) and Eq.(\ref{z4}) respectively. 

There are a number of vertex excitations created by these operators and hence it is possible to write down a similar table to table(\ref{sometable}). Instead we will just show the presence of vertex excitations that mimic those of the quantum double of $S_3$. It is known that the non-Abelian charges excitations of the quantum double of $S_3$ are created by~\cite{bombinExcite, LevinNA},
\begin{eqnarray} F_{\rho ; B(e)}^{(1,1)(1,1)} & = & \frac{1}{3}\left(\begin{array}{cc} Z^2 & 0 \\ 0 & 0\end{array}\right) \\
F_{\rho ; B(e)}^{(1,2)(1,2)} & = & \frac{1}{3}\left(\begin{array}{cc} Z & 0 \\ 0 & 0\end{array}\right) \\
F_{\rho ; B(e)}^{(1,1)(1,2)} & = & \frac{1}{3}\left(\begin{array}{cc} 0 & 0 \\ 0 & Z\end{array}\right) \\
F_{\rho ; B(e)}^{(1,2)(1,1)} & = & \frac{1}{3}\left(\begin{array}{cc} 0 & 0 \\ 0 & Z^2\end{array}\right). \end{eqnarray}

The notation used is here is the standard notation for ribbon operators as found in~\cite{bombinExcite} and so will not be explained here. 

Clearly these operators are part of Eq.(\ref{excitations63}) and hence are part of the vertex excitations of the $(6,3)$ model. 

We again have a local operator acting on the vertex which commutes with both the vertex and the link operators.This is given by 
 \beq R_v = \left(\begin{array}{cc} a\mathbb{1} & 0 \\ 0 & b\mathbb{1}\end{array}\right)\eeq 
where $a$ and $b$ are arbitrary constants and $\mathbb{1}$ is the three by three identity matrix.

Having obtained the excitation operators in the three examples considered we will now study the fusion rules of these systems and show they obey those of non-Abelian systems. 

\section{Non-Abelian Fusion Rules}

The fusion rules of a given set of anyons can be computed by looking at the algebra of the operators creating those excitations. For example in the $\mathbb{Z}_2$-toric code case we have 4 types of particles given by $(1, e, m, \epsilon)$. The operators creating these excitations in this case are given by $\mathbb{1}, \sigma^z, \sigma^x, \sigma^y$ respectively. The fusion rules for these excitations given by $e\times e = m\times m= \epsilon\times \epsilon=1$, $e\times m=\epsilon$ and $e\times 1 = 1\times e=e,  m\times 1= 1\times m=m, \epsilon\times 1= 1\times \epsilon =\epsilon$ follows from the algebra of the two by two Pauli matrices which create these excitations. In a similar fashion we can find the algebra of the operators creating excitations in the $H_n/\mathbb{C}(\mathbb{Z}_p)$ models to find their fusion rules. We now show this for the three examples considered.

\subsection{$(n,p) = (3,2)$:}

The vertex excitations in the $(3,2)$ case are given by $(X_{12}, Y_{12}, Q_+^x, Q_-^x, Q_+^y, Q_-^y, Q_+^z, Q_-^z)$. These are shown in table(\ref{sometable}). The fusion rules are given by
\bea Q_{\mp}^x\times Q_{\mp}^y & = & iQ_+^z \mp iX_{12};~ Q_{\mp}^x\times Q_{\pm}^y  =  -iQ_-^z \pm Y_{12} \nn \\
Q_+^y\times Q_-^y & = & -Q_-^z + iY_{12};~ Q_+^x\times Q_-^x = -Q_-^z+iY_{12}\nn \\
Q_+^z\times Q_+^x & = & - \frac{1}{2}Q_+^x + \frac{3i}{2}Q_+^y;~ Q_+^z\times Q_-^x = -\frac{1}{2}Q_-^x - iQ_+^y -\frac{i}{2}Q_-^y \nn \\ 
Q_+^z\times Q_+^y & = & \frac{3i}{2} Q_+^x + \frac{1}{2}Q_+^y;~ Q_+^z\times Q_-^y = - \frac{3i}{2}Q_-^x - Q_+^y - \frac{3}{2}Q_-^y \nn \\
Q_+^z\times Q_-^z & = & Q_-^z;~ Q_+^z\times X_{12} = X_{12};~ Q_-^z\times Q_-^z = \frac{2}{3}1 + \frac{1}{3}Q_+^z   \nn \\ 
Q_-^z\times Q_{\pm}^x & = & - \frac{1}{2}Q_{\mp}^x - \frac{i}{2}Q_{\mp}^y;~ Q_-^z\times Q_{\pm}^y = \frac{i}{2}Q_{\mp}^x - \frac{1}{2}Q_{\mp}^y \nn \\
X_{12}\times Q_{\pm}^x & = & \pm\frac{1}{2}Q_{\pm}^x \pm \frac{i}{2}Q_{\pm}^y;~ X_{12}\times Q_{\pm}^y  =  \mp \frac{i}{2} Q_{\pm}^x \pm \frac{1}{2} Q_{\pm}^y  \nn \\ 
 Q_+^z\times Q_+^z & = & 2 - Q_+^z;~Q_{\pm}^x\times Q_{\pm}^x = \frac{4}{3}{\pm}X_{12}-\frac{1}{3}Q_+^z \nn \\ 
Q_{\pm}^y\times Q_{\pm}^y & = & \frac{4}{3}{\pm}X_{12}-\frac{1}{3}Q_+^z;~Y_{12}\times Q_+^z = Y_{12} \nn \\ 
Y_{12}\times Q_{\pm}^x & = & \pm \frac{i}{2}Q_{\mp}^x \mp \frac{1}{2}Q_{\mp}^y;~ Y_{12}\times Q_{\pm}^y = \pm \frac{1}{2}Q_{\mp}^x \pm\frac{i}{2}Q_{\mp}^y \nn \\ 
X_{12}\times Y_{12} & = & iQ_-^z;~ X_{12}\times Q_+^z = X_{12} \nn \\ 
Q_-^z\times X_{12} & = & iY_{12};~ Q_-^z\times Y_{12} = -iX_{12}. \eea

 These fusion rules are clearly non-Abelian. However we are not sure what anyons these are, that is if they come from the irreducible representations of any quantum double. They can nevertheless be thought of as superpositions of the excitations of a quantum double model based on $\mathbb{Z}_3$. 

\subsection{$(n,p)=(4,2)$:}

We now identify the fusion rules of Ising anyons in the $(4,2)$ case. The operator in Eq.(\ref{ising}) is the non-Abelian particle of the Ising anyons. We show this by looking at their fusion rules which are
\beq \Sigma_k\times\Sigma_m = \left(km+1\right)\left(\begin{array}{cc} 1 & 0 \\ 0 & 1\end{array}\right) + i\left(k-m\right)\left(\begin{array}{cc} \sigma^y & 0 \\ 0 & \sigma^y\end{array}\right) \eeq
where $k$ and $m$ are arbitrary parameters. 

The first operator in this rule is the vacuum of the Ising anyon particle set. The second one corresponds to the fermion $\psi$ as denoted in the standard notation for Ising anyons~\cite{PachosIsing}. So for $k=m=1$ we obtain the vacuum sector and for $k=-m=1$ we obtain the fermion $\psi$. Note that $\left(\begin{array}{cc} \sigma^x & 0 \\ 0 & \sigma^x\end{array}\right)$ creates a link excitation and commutes with the vertex operator. The other part of $\Sigma_k$ given by $\left(\begin{array}{cc} \sigma^z & 0 \\ 0 & \sigma^z\end{array}\right)$ creates a vertex excitation but commutes with the link operator. And finally $\left(\begin{array}{cc} \sigma^y & 0 \\ 0 & \sigma^y\end{array}\right)$ creates an excitation for both the vertex and link operators.  

\subsection{$(n,p) = (6,3)$:}

For this case again we do not consider all possible excitations as there are too many of them. As we had mentioned in the previous section, where we studied the operators creating excitations, the operators creating excitations in the quantum double model based on $S_3$ are part of the operators creating excitations in the $(6,3)$ model. Thus we can be sure that these obey the fusion rules of $S_3$ anyons.  

\section{Remarks}

We have exhibited a class of lattice models with gauge and matter degrees of freedom in two dimensions. We studied the vertex excitations in these models and showed that they obey non-Abelian fusion rules for $n>p$. This was seen through three examples namely $(3,2)$, $(4,2)$ and $(6,3)$. The latter two were shown to host the familiar Ising anyons and the anyons of the $S_3$ quantum double model. These models are thus examples of systems which are not quantum double models but host non-Abelian anyons as part of their low energy excitations. 

These models were originally constructed from transfer matrices built using a procedure similar to the one used in~\cite{QDMST}. This construction is quiet elaborate and needs more space and so we will present this is in a separate paper~\cite{PPPM}. In other words these can be thought of as generalizations of~\cite{QDMST} where we use Kuperberg invariants to build transfer matrices for lattice gauge theories based on involutary Hopf algebras. In particular the Hamiltonians constructed out of these transfer matrices contained the quantum double Hamiltonians for particular choices of parameters~\cite{QDMST}. The advantage of this formalism is that it allows us to study a wide range of models in the parameter space and thus helps understand phase transitions between topologically ordered and non-topological phases. A similar study can be carried out for theories with gauge and matter fields~\cite{PPPM}. This helps us obtain a variety of models. In particular we can obtain Kitaev's honeycomb models when we consider $H_2/\mathbb{C}(\mathbb{Z}_2)$ as our input.

The models presented in this paper can be thought of as conceptual generalizations of the Kitaev Honeycomb model~\cite{KitHoney} which are known to host Ising anyons when magnetic fields are added to the system. The non-Abelian phase of the Kitaev Honeycomb model is suitable for quantum computation as the braid group, acting on the excitations of the low energy sector in the theory, is not superselected. That is it does not commute with the operators creating the excitations in the theory. This is crucial to realize quantum gates using the non-Abelian anyons as now we can experimentally probe the statistics of these particles. The $H_n/\mathbb{C}(\mathbb{Z}_p)$ model also has this feature and thus maybe suitable for quantum computation. We plan to explore this aspect of this problem in a future work. 

We believe that for all $n>p$ these models host non-Abelian vertex excitations. It will be interesting to find appropriate $n$ and $p$ for a given set of anyonic fusion rules.

We did not study the braiding of these excitations. This is however easy to achieve as we have the operators creating the excitations and thus they can be used to move these particles.   

Lattice models for topological insulators can be constructed using a procedure similar to~\cite{QDMST} for appropriate choice of gauge groups and their modules. This formalism will thus help in studying the parameter space of theories with gauge and matter fields and so can possibly help in studying transitions between normal and topological insulators. We will explore these problems in future works. 

\section*{Acknowledgements}

PP thanks A.P.Balachandran for useful discussions especially the interpretation of the non-Abelian sector of the Kitaev Honeycomb model. The authors thank Pablo Ibieta for help in preparing the manuscript. The authors thank FAPESP for support during this work.

\end{document}